%% file: main.tex
\documentclass[sigconf, anonymous=False, review=False]{acmart}

\usepackage{microtype}
\usepackage{graphicx}
\usepackage{subfigure}
\usepackage{enumitem}
\usepackage{booktabs} 
\usepackage{bm}
\usepackage{times}
\usepackage{graphicx} 
\usepackage{subfigure} 
\usepackage{amsmath}
\usepackage{amsfonts}
\usepackage{mathtools}
\usepackage{ctable}
\usepackage{multirow}
\usepackage{stmaryrd}
\usepackage{float}

\usepackage{makecell}

\usepackage{natbib}

\usepackage{algorithm}
\usepackage{algorithmic}

\usepackage{hyperref}

\AtBeginDocument{%
  }

\setcopyright{acmcopyright}
\copyrightyear{2022}
\acmYear{2022}
\acmDOI{XXXXXXX.XXXXXXX}

\acmConference[Conference acronym 'XX]{DLP-KDD}{2022}{Washington, DC}
\acmPrice{15.00}
\acmISBN{978-1-4503-XXXX-X/18/06}

\begin{document}

\title{Ask Me What You Need: \\ Product Retrieval using Knowledge from GPT-3}

\author{Su Young Kim$^{1}$, Hyeonjin Park$^{1}$, Kyuyong Shin$^{1,2}$, Kyung-Min Kim$^{1,2}$}
\email{{suyoung.kim1, hyeonjin.park.ml, ky.shin, kyungmin.kim.ml}@navercorp.com}
\affiliation{%
  \institution{$^{1}$ NAVER CLOVA, $^{1,2}$ NAVER AI Lab}
  \country{South Korea}
  }
\renewcommand{\shortauthors}{Trovato et al.}

\begin{abstract}
As online merchandise become more common, many studies focus on embedding-based methods where queries and products are represented in the semantic space. These methods alleviate the problem of vocab mismatch between the language of queries and products. However, past studies usually dealt with queries that precisely describe the product, and there still exists the need to answer imprecise queries that may require common sense knowledge,
\textit{i.e.}, `what should I get my mom for mother's day.' 
In this paper, we propose a GPT-3 based product retrieval system that leverages the knowledge-base (KB) of GPT-3 for question answering; users do not need to know the specific illustrative keywords for a product when querying. 
Our method tunes prompt tokens of GPT-3 to prompt knowledge and render answers that are mapped directly to products without further processing. 
Our method shows consistent performance improvement on two real-world and one public dataset, compared to the baseline methods.
We provide an in-depth discussion on leveraging GPT-3 knowledge into a question answering based retrieval system.

\end{abstract}




\maketitle

\section{Introduction}
\input{01.Introduction}
\section{Methodology}
\input{03.Model}
\section{Experiments}
\input{04.Experiments}
\section{Results}\label{result_sec}
\input{05.ComparsionResults}
\section{Conclusion}
\input{06.Conclusion}
\bibliographystyle{ACM-Reference-Format}
\bibliography{reference}
\end{document}

%% file: 01.Introduction.tex
Product search engines have emerged as a key factor for online e-commerce platforms. They allow users to find the best set of products offered by an online merchandise that match the search query.
Early work investigating product retrieval focused on improving a lexical matching engine that quantifies the similarity between languages of query and product ~\cite{ huang_sharma_sun_xia_zhang_pronin_padmanabhan_ottaviano_yang_2020, 10.1145/3292500.3330759}.
However, such methods often suffered from an upper bound in the semantic information they can learn. 
More recent studies introduced neural product search approaches, where latent representations for queries and products are modelled using a deep neural network \cite{10.1145/3404835.3462911, 10.1145/3459637.3482358}. 
State-of-the-art product retrieval models utilize pre-trained large language models to enhance text embedding and further bridge the vocabulary gap.  \citet{10.1145/3485447.3511977} modelled query and product representations using fine-tuned BERT, and yield an end-to-end learning framework for product search. \citet{https://doi.org/10.48550/arxiv.2105.02978} used a BERT-based query encoder and a graph attention based retrieval network for e-commerce search.

In this paper, we deal with a particular challenge of modelling \textit{intent} queries to retrieve relevant products. These intent queries are different from keyword queries that explicitly describe the desired products. Users may not know the relevant item but only know the search intent, \textit{i.e.}, `What should I get my son for his birthday?' Inspired by the properties of recent large language models, we expect to obtain such extra common sense knowledge from GPT-3.

 

GPT-3 has shown great success in Natural Language Processing (NLP) domains such as question answering (QA)~\cite{NEURIPS2020_1457c0d6, kim2021changes, nakano2021webgpt} and knowledge retrieval~\cite{yang2021empirical, nakano2021webgpt}.
\input{Figures_tex/0.main_figure}
Instead of requiring an explicit knowledge base (KB), our product retrieval engine uses an \textit{implicit} KB that is stored in GPT-3, along with p-tuning~\cite{liu2021gpt} to better prompt the stored knowledge. 
Our main contributions are summarized as follows. 
\begin{enumerate}[leftmargin=*]
    \item We present a GPT-3 based product retrieval system. To the best of our knowledge, this is the first use of GPT-3 in a product retrieval task.
    \item We conduct ablation studies on how GPT-3 size and tuning methods affect the performance of the product retrieval system.
    \item We show that GPT-3 based product retrieval system is more effective in solving the cold-start problem than other baselines.
\end{enumerate}

%% file: Figures_tex/0.main_figure.tex
\begin{figure*}[ht] 
\centering
 \includegraphics[width=0.9\textwidth]{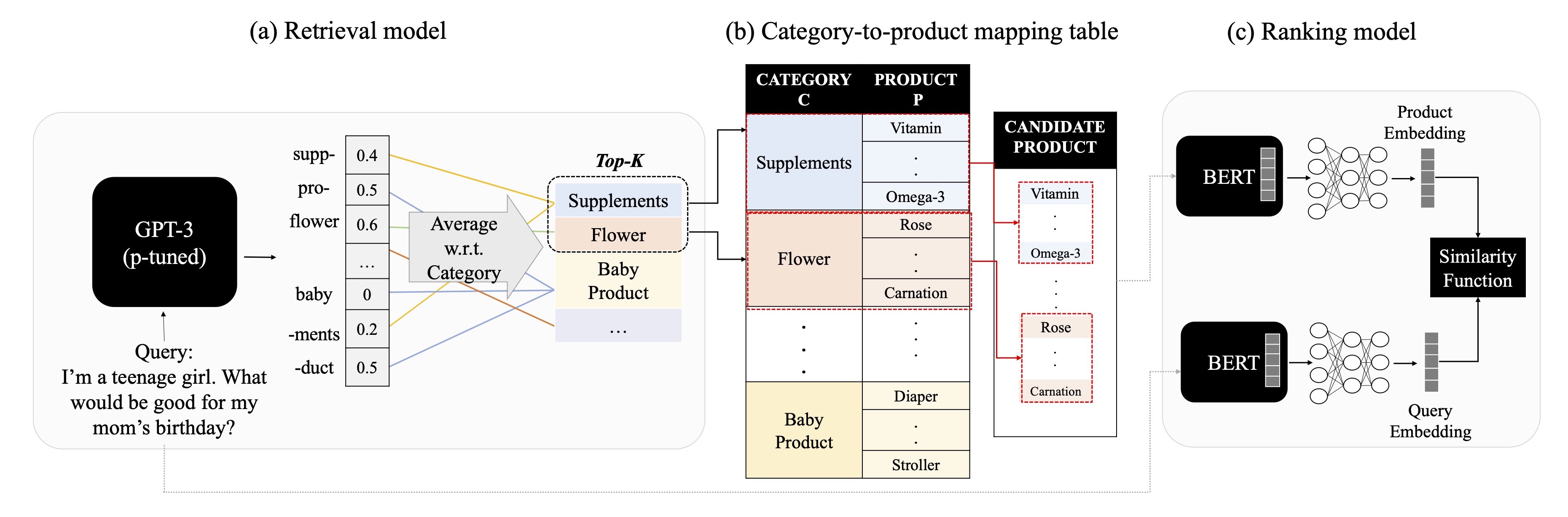}
 \caption{The overview of the proposed product retrieval system. (a) The retrieval model takes a query and selects top-$K$ categories according to the category score. (b) The top-$K$ categories are mapped to candidate products using category-to-product mapping table. (c) Lastly, the candidate products are ranked according to the ranking model.}
\label{fig:main_fig}
\end{figure*}

%% file: 03.Model.tex
First, we define the notations used in this paper. Let $Q = \{q_1, \ldots, q_n\}$ be a set of $n$ queries and $C = \{c_1, \ldots, c_m\}$ be a set of $m$ categories. Each query $q_i$ has a purchased product $p_j \in P$ and the product's corresponding categorical information $c_k \in C$. The triplet of $q_i,p_j$ and $c_k$ appear in the data logs, \textit{i.e.}, a user had a search intent $q_i$ and purchased $p_j$ belonging in category $c_k$. 

\subsection{Retrieval model}
Given a query $\hat{q}_i$, the goal of our retrieval model is to select the top-$K$ relevant product categories $\hat{C}$, thus \textit{effectively} reducing the search space for the subsequent ranking model.
We choose products' categorical information as the target label for the retrieval stage because using categories as answers makes the template closer to natural language as a human would write it, and the performance of GPT-3 would improve.

\vspace{0.05in}
\noindent\textbf{Training method.}
To optimize GPT-3 for our downstream task, we use the p-tuning method~\cite{liu2021gpt}. 
We formulate $\tilde{q}_i$ as a concatenation, "[PROMPT$_{1:d}$] [$\hat{q}_i$] [MASK]", in which [PROMPT$_{1:d}$] are the trainable continuous prompt tokens, [$\hat{q}_i$] is the context, and [MASK] is the target. $d$ is the hyperparameter determining the number of prompt tokens. In p-tuning method, only the embeddings for the trainable continuous prompt tokens are updated with the Cross-Entropy loss,
\begin{equation}
        L = - \sum_{i=1}^{N}y_i^{\top}log\mathcal{M}(\tilde{q}_i),
\end{equation}
where $\mathcal{M}$ refers to the GPT-3 model, $N$ is the number of train data and $y_i$ is the one-hot vector of the target category token in the vocabulary of a language model $\mathcal{M}$.
In practice, fine-tuning could also be adopted, but many work observed that GPT-style models perform poorly to NLU tasks with fine-tuning \cite{liu2021gpt}. 
Thus, we use p-tuning in utilizing GPT-3 as an implicit KB for desired knowledge, in other words, finding a relevant product's category for a given query. The comparison between performances of fine-tuning and p-tuning is discussed in Section~\ref{sec_tuning}.

\vspace{0.05in}
\noindent\textbf{Inference.}
To select the top-$K$ relevant product categories $\hat{C}$, we obtain the category score $s_i$ for each category. Let's say the category $c_i$ is `baby product' and its tokens are $T = [\text{`baby'}, \text{`pro-'}, \text{`-duct'}]$.
$s_i$ is formulated as,
\begin{equation}
        s_i = \sum_{j=1}^{|T|}{\alpha_{j}\mathcal{M}(t_j | \tilde{q})},
\end{equation}
where $t_j$ denotes $j$-th token in $T$. 
The $\alpha_j$ is the hyperparameter for the weight of each token probability and $\sum_j \alpha_j=1$. 
$s_i$ is calculated as the weighted average of the logit scores of tokens in $c_{i} \in C$, conditioned on the query.

In our experiment, we heuristically set $\alpha_{1}$ to $0.8$ and let the rest share the same weight of $\sum_{j=2}^{|T|}\alpha_j=0.2$. We give the highest weight to the first token because it is the most important in decoding the answer from GPT-3. Finally, category set $C$ is sorted by $s_i$ and the top-$K$ categories $\hat{C}$ are used in the ranking stage.

\subsection{Ranking Model}
We first use the category-to-product mapping table, ~\figureautorefname~\ref{fig:main_fig}-(b), to prepare the candidate product set.
The candidate products are then ranked using the ranking model, which can be any model that leverages embedding-based similarity methods. 

In this paper, we use BERT~\cite{devlin2019bert} with multi-layer perceptron (MLP) layers as the simple embedding method to leverage flexibility in the architecture. Learning latent representations of queries and products with this embedding method and then calculating a similarity score is shown in ~\figureautorefname~\ref{fig:main_fig}-(c). 
Specifically, given the query $\hat{q}$ and the candidate product $\hat{p_i}$, the similarity score is calculated as,
\begin{equation}
    \begin{aligned}
    S(\hat{q}, \hat{p}_i) = f(E_{\hat{q}}; \theta) \cdot  f(E_{\hat{p}_i}; w),
    \end{aligned}
\end{equation}
where $E_{\hat{q}}$ and $E_{\hat{p}_i}$ represent the BERT embeddings of $\hat{q}$ and $\hat{p}_i$, and $f$ is MLP layers parameterized by $\theta$ and $w$. 

We use Binary Cross-Entropy (BCE) loss between the predicted score $S(\hat{q},\hat{p}_i)$ and ground truth label $\hat{y}$ which is defined as, 
\begin{equation}
    \hat{y} =     
    \begin{cases}
    1, & (\hat{q}, \hat{p}_j) \in D^{tr}\\
    0, & \textrm{otherwise}
    \end{cases}
\end{equation}
where $D^{tr}$ is the set of query and product pairs $(q_i, p_j)$ in the data log.
We also use the weighted loss to handle the class imbalance problem between positive and negative samples. The weight update happens both in the BERT parameters and MLP parameters.

%% file: 04.Experiments.tex
\input{tab_dataset}

\subsection{Datasets and Evaluation Metrics}
In this section, we conduct experiments on our e-commerce datasets, where the purchase log is transformed into query-product pairs. These queries do not describe the products, but instead, contain search intents that require the system to exhibit natural language understanding (NLU) ability. We additionally test our method on a public dataset. 
Detailed statistics of the datasets are described in~\tableautorefname~\ref{tab:dataset}.

\vspace{0.05in}
\noindent{\textbf{Gift dataset.}} We retrieved reviews that contain the word `gift' from one year of shopping review log on our e-commerce platform, spanning from May 20, 2020, to May 25, 2021. 
We subsampled a total of $55,217$ review logs, then involved human resources to form natural language queries from the review logs, to produce query-product pairs. 
Since the log contains user information, we could compute Toppop (age or gender) for users asking the queries, as baselines.

\vspace{0.05in}
\noindent{\textbf{Co-purchase dataset.}} We sampled $45,234$ purchase logs from our e-commerce platform, spanning from September 01, 2021, to September 5, 2021.
For each purchase log, we randomly picked a product as an anchor and formed a query with its category information, \textit{i.e.}, if the anchor product is a type of vitamin, the query is `What can be co-purchased with vitamins?'
The positive sample is a randomly picked product from the same purchase log. Additionally, to formulate it into a more complex NLU task, we asked 82B GPT-3~\cite{kim2021changes} for the intention of the co-purchase to include it in the query. 
\footnote{We exclude TopPop for Co-purchase since this dataset is collected
without demographic information}

\vspace{0.05in}
\noindent{\textbf{Google LCC}}
To verify the generality of our proposed method and allow others to reproduce our results, we additionally test on a public dataset shared by Google LCC.
This dataset consists of $6,555$ English question-answer pairs, which are categorized as one of `stackoverflow', `culture', `technology', `science', or `life arts' based on the nature of the question. In this paper, we viewed questions as queries, answers as products, and categories as product categories.

\vspace{0.05in}
\noindent{\textbf{Evaluation metrics.}}
We evaluate model performance on Gift,  Co-purchase and Google LCC dataset in terms of HR@$K$, a simple yes/no metric that looks at whether any of the top-$K$ recommended products include the ground truth product,
\begin{equation}
    \textrm{HR}@K \textrm{ for a query} = \max_{i=1,..,K}
    \begin{cases}
    1, & r \in \mathcal{T}_i\\
    0, & \textrm{otherwise}
    \end{cases}
\end{equation}
where $r$ is the ground truth product, and $\mathcal{T}_i$ is the set of recommended products up to $i$-th rank. We get HR@$K$ across queries and compute the average.



\subsection{Methods Compared}
We compare against a conventional baseline (TopPop), a traditional web retrieval baseline (BM25~\cite{thakur2021beir}), and an transformer-based baseline that is widely used for NLP modelling (BERT~\cite{devlin2019bert}). All these baselines are formed as a 2-stage retrieval system, where the retrieval model follows each baseline method but the same ranking model as ours is shared across all baselines. Note that the top 10 categories were retrieved for the gift and co-purchase dataset, whereas the top one category was retrieved for the Google LLC dataset.

\vspace{0.05in}
\noindent{\textbf{TopPop.}} Toppop baseline retrieves categories according to the category popularity. We test toppop on two levels, age and gender of the user asking the query. 

\vspace{0.05in}
\noindent{\textbf{BM25.}} Overall, BM25 remains a strong baseline for zero-shot text retrieval~\cite{thakur2021beir}. BM25 is a bag-of-words (BOW) information retrieval model that relies on an exact lexical match between a query and documents (categories).

\vspace{0.05in}
\noindent{\textbf{BERT-based similarity search.}} 
The current effective approaches integrate BERT~\cite{devlin2019bert} as an embedding generation component in the retrieval model, with the input of a concatenated string of query and candidate texts. BERT and a simple nonlinear model are then trained with BCE loss where incorrect pairs get penalized. 
\input{tab_all}
\input{tab_information_retrieval}

%% file: tab_dataset.tex
\setlength{\tabcolsep}{3pt}
\ctable[
    caption = {Basic Statistics of Datasets.
},
    label = tab:dataset,
    pos=t,
 	doinside=\normalsize
]{cccccc}{
}{

\toprule \multirow{2}{*}{Dataset}  & \multicolumn{3}{c}{\# of Pair} & \multirow{2}{*}{\# of Categories} & \multirow{2}{*}{\# of Items} \\
\cmidrule(l){2-4}   &   Train & Valid & Test &  & \\
\midrule[0.52pt]
\midrule[0.52pt]
     Gift & 44,173 & 5,522 & 5,522 & 1,357 & 41,589 \\
     \midrule
     Co-purchase & 67,524 &8,441& 8,440 & 1,076 & 64,020 \\
     \midrule
     Google LCC & 5603 & 476 & 476 & 5 & 6555\\
\bottomrule
}

%% file: tab_all.tex
\setlength{\tabcolsep}{4pt}
\ctable[
    caption = {Results on the two product retrieval datasets.},
    label = tab:all,
    pos=t,
 	doinside=\normalsize
]{ccccc}{
}{
\toprule
   \multirow{2}{*}{Retrieval Models} & \multicolumn{2}{c}{Gift dataset} & \multicolumn{2}{c}{Co-purchase dataset} \\
   \cmidrule(l){2-3}
   \cmidrule(l){4-5}
   & HR@$300$ & HR@$500$  & HR@$300$ & HR@$500$ \\
\midrule
\midrule
TopPop (age)  & 0.0541 & 0.0630 & - & -  \\
TopPop (gender) & 0.0590  & 0.0670 & - & - \\
\midrule
BM25 & 0.0060 & 0.0063 & 0.0065 & 0.0075 \\
BERT  & 0.1317 & 0.1492  & 0.1135 & 0.1609 \\
\midrule
Ours (13B) &  \textbf{0.1514} & \textbf{0.1699}  & \textbf{0.1494} & \textbf{0.2121} \\
\bottomrule
}

%% file: tab_information_retrieval.tex
\setlength{\tabcolsep}{4pt}
\ctable[
    caption = { Results the Google LCC dataset.},
    pos=t,
 	doinside=\normalsize
]{ccccc}{
}{
\toprule
   \multirow{2}{*}{Retrieval Models} & \multicolumn{2}{c}{Google LCC dataset}  \\
  \cmidrule(l){2-3}
   & HR@$300$ & HR@$500$\\

\midrule
BM25 & 0.0945 & 0.1176 \\
BERT  & 0.0840 & 0.1450  \\
\midrule
Ours (6.7B) &  \textbf{0.3004} & \textbf{0.4685} \\
\bottomrule
}
\vspace{0.05in}

%% file: 05.ComparsionResults.tex
\subsection{Retrieval Performance}\label{sec_main_result}
\subsubsection{Product Retrieval}
We have our main comparison table in ~\tableautorefname~\ref{tab:all}. We see superior performance of our proposed method on both datasets, across all metrics. BM25~\cite{thakur2021beir} retrieval model showed the worst performance, as the tasks are formed as QA-tasks, therefore the lexical matching method is significantly suboptimal. TopPop baselines also failed to retrieve relevant categories to the query. BERT~\cite{devlin2019bert}-based similarity search was the most comparable considering it is a transformer-based pre-trained language model, however, our proposed retrieval system showed superior performance of 15.0\% and 31.6\% compared to BERT. The higher performance of our method comes from the ability to carefully consider the scores of each token of the category.
\subsubsection{Public Dataset}
Our method on Google LLC dataset showed superior performance compared to BERT~\cite{devlin2019bert} and BM25~\cite{thakur2021beir}, proving that our method can be generalized to not only product retrieval but also to general informational retrieval cases. The more conspicuous performance increase results from the nature of the dataset. The product retrieval dataset has many possible answers for one ground truth product, whereas in the Question-Answering scenario the right answer is apparent, thus the ranking model could perform better.

\input{tab_tuning}

\input{tab_cold}

\subsection{Effective Tuning Method for Knowledge Retrieval}\label{sec_tuning}
Interestingly, fine-tuning GPT-3 shows only a slight performance improvement compared to the zero-shot approach (\tableautorefname~\ref{tab:tuning}). We conjecture that fine-tuning the model parameters triggered catastrophic forgetting, which subsided the knowledge GPT-3 gained from pre-training.
\citet{liu2021gpt} empirically demonstrated that pre-trained language models with properly optimized p-tuning can capture far more knowledge than fine-tuning, and show that such is true across various model scales for NLU tasks. We observe the same trend in our KB-based product retrieval system. ~\tableautorefname~\ref{tab:tuning} shows that our proposed method with $137$ million parameters significantly outperforms the other tuning methods, which are  $1.3$ billion zero-shot and $1.3$ billion fine-tuned models.

\subsection{Influence of Language Model Size}\label{sec_LM_size}
Recent studies have shown that training deep neural networks with large parameters leads to significant performance gains in a wide range of tasks~\cite{NEURIPS2020_1457c0d6, shin2021scaling}. As such, we found that scaling up the size of GPT-3 empowers the ability to solve QA tasks. As presented in~\tableautorefname~\ref{tab:tuning}, 13 billion model surpasses other models by a very great margin. It is worth noting that the performance differs even more significantly when the model size varies from 1.3 billion to 13 billion, than from 137 million to 1.3 billion. This implies that increasing the model size can dramatically increase the knowledge probing ability of the language model.

\subsection{Performance on Cold-Start Problem}\label{sec_cold}
We evaluate the cold-start performance against two other baselines, BM25 and BERT-based search. To properly compare the performance, we take the same train dataset as before but prepare a separate test dataset consisting of search intent and product pairs not seen during training. Our method achieves an increase of 85.8\% in HR@300 metric compared to BERT-based search (\tableautorefname~\ref{tab:cold}). We conclude that the knowledge already encoded in GPT-3 helps retrieve the right products, although a particular query or product's semantic information has not been learned during training. Our product retrieval system overcomes the vocabulary upper bound problem as well as allowing flexibility in query formation. 



%% file: tab_tuning.tex
\setlength{\tabcolsep}{18pt}
\ctable[
    caption = {Performance comparison of different tuning methods on Gift dataset.},
    label = tab:tuning,
    pos=!tbp,
 	doinside=\normalsize
]{ccc}{
\tnote[$\ast$] {GPT-3 (13B) is too large for fine-tuning.}
}{
\toprule
   Retrieval Models & HR@$300$ & HR@$500$ \\
\midrule
Zero-shot (1.3B) & 0.0076 & 0.0130 \\
Fine-tuned (1.3B) & 0.0092 & 0.0158\\
\midrule
Ours (137M)  & 0.0570 & 0.0954 \\
Ours (1.3B) & 0.0628 & 0.0996 \\
Ours (13B) & \textbf{0.1514} & \textbf{0.1699} \\
\bottomrule
}

%% file: tab_cold.tex
\setlength{\tabcolsep}{18pt}
\ctable[
    caption = {The cold-start performance on Gift dataset.},
    label = tab:cold,
    pos=t,
 	doinside=\normalsize
]{ccc}{
}{
\toprule
   Retrieval Models & HR@$300$ & HR@$500$ \\
\midrule
BM25  & 0.0002 & 0.0009  \\
BERT & 0.0141 & 0.0235 \\
\midrule
Ours (13B) & \textbf{0.0262} & \textbf{0.0410} \\
\bottomrule
}

%% file: 06.Conclusion.tex
We propose a GPT-3 based product retrieval system that can leverage the implicit KB stored in GPT-3 to answer intent queries that may contain out-distribution vocabularies. 
Our method is non-trivial because it uses GPT-3 in product retrieval tasks while using p-tuning to guide the prompting of knowledge and retrieval of information. We test our method on two real-world and one public dataset, where we see superior performance compared to the competitive baseline, BERT, even in the cold-start setting.
In the future, we plan to develop a personalized product retrieval system that integrates individual user behavior logs into our system.